\documentclass[aps,prl,twocolumn,superscriptaddress,reprint]{revtex4-2}
\usepackage{float}
\usepackage{graphicx}
\usepackage{amsmath}
\usepackage{bm}
\usepackage{color}
\usepackage{amssymb}
\usepackage{mathrsfs}
\usepackage{dcolumn}
\usepackage{multirow}
\usepackage[mathlines]{lineno}
\usepackage[colorlinks=true,linkcolor=blue,urlcolor=blue,citecolor=blue]{hyperref}

\begin{document}

\title{Emergent Weyl Fermions in an Orbital Multipolar Ordering Phase} 

\author{Hua Chen}
\affiliation{Department of Physics, Zhejiang Normal University, Jinhua 321004, China}

\author{Congjun Wu}
\email{Electronic address: wucongjun@westlake.edu.cn} 
\affiliation{School of Science, Westlake University, Hangzhou 310024, China}
\affiliation{Institute for Theoretical Sciences, Westlake University, Hangzhou 310024, Zhejiang, China}
\affiliation{Key Laboratory for Quantum Materials of Zhejiang Province, School of Science, Westlake University, Hangzhou 310024, China}

\author{X. C. Xie}
\affiliation{International Center for Quantum Materials, School of Physics, Peking University, Beijing 100871, China}
\affiliation{CAS Center for Excellence in Topological Quantum Computation, University of Chinese Academy of Sciences, Beijing 100190, China}	
	
\begin{abstract}
Multipolar orderings in degenerate orbital systems offer unique opportunities for emergent topological phases. The phase diagram of interacting spinless fermions in a $p$-band diamond lattice at unit filling is first studied to elucidate the essential role of orbital multipolar orderings in the evolution of multifold degenerate band nodes. The free band structure around the Brillouin zone center is described by two quadratic band nodes each with a threefold degeneracy, which are spanned by the bonding and anti-bonding $p$-orbital multiplets, respectively. Upon switching on interactions, the triply degenerate band node is split into a pair of Weyl fermions with opposite chirality due to the onset of orbital multipolar orderings. Further raising interactions ultimately drives the system into an insulating phase with the orbital quadrupolar ordering. Our study is then generalized to spin-$1/2$ fermions, which has direct relevance with solid-state materials. The system develops full spin polarization through a ferromagnetic transition at tiny interactions, leaving the remaining orbital sector activated. The ensuing transitions take place in the orbital sector as a natural consequence, qualitatively recovering the phase diagram of spinless fermions. Our findings shed new light on the realization of emergent novel fermions with a prospect being a frontier at the confluence of topology, orbital physics and strong correlation.
\end{abstract}

\date{\today}
	
\maketitle

\begin{table}[t]
	\centering 
	\begin{tabular}{c|c|c|c} 
		\hline
		Multipole 				&	Time Reversal  		& IR 				& Multipole Operator $\bm{X}$ \\ 
		\hline \hline 
		Monopole		&	Even		& $\Gamma_1^+$			& $\hat{n}=\frac{1}{2\sqrt{3}}\bm{l}^2$		\\
		\hline
		Dipole 	&	Odd	& $\Gamma_{15}^+ $ 			& $\hat{J}_x=\frac{1}{\sqrt{2}}l_x$  	\\  
		&      			&					& $\hat{J}_y=\frac{1}{\sqrt{2}}l_y$ 	\\
		&          		&					& $\hat{J}_z=\frac{1}{\sqrt{2}}l_z$ 	\\
		\hline
		Quadrupole 	& Even	& $\Gamma_{25}^+$	&$\hat{O}_{yz} = \frac{1}{\sqrt{2}} \{l_y,l_z\}$ 	\\
		&		&					&$\hat{O}_{zx} = \frac{1}{\sqrt{2}} \{l_z,l_x\}$  \\
		&		&					&$\hat{O}_{xy} = \frac{1}{\sqrt{2}} \{l_x,l_y\}$ 	\\
		&		& $\Gamma_{12}^+$	& $\hat{O}_{x^2-y^2} = \frac{1}{\sqrt{2}}(l^2_x- l^2_y) $  	\\
		&		&					& $\hat{O}_{3z^2-r^2} =\frac{1}{\sqrt{6}} (3l_z^2- \bm{l}^2)$\\ 
		\hline                              
	\end{tabular}
	\caption{
		The decomposition of $p$-orbital density matrix into multipole operators $\bm{X}$ with effective angular momenta $\ell=0,1,2$, denoted by monopole (charge) $\hat{n}$, dipole $\bm{\hat{J}}$, and quadrupole $\bm{\hat{O}}$ respectively, according to the irreducible representations (IRs) of cubic group symmetry $O_h$. Here $l_\mu$ is the matrix form of angular momentum operator $\hat{l}_\mu=-i\sum_{\mu\nu\gamma}\epsilon_{\mu\nu\gamma}p^\dagger_\nu p_\gamma$ in $p$ orbital bases with $\epsilon_{\mu\nu\gamma}$ being the Levi-Civita antisymmetric tensor. The parity of time reversal symmetry is also classified. The multipole operators satisfy the orthonormality condition $\text{Tr}\left[X_\alpha^\dagger X_\beta\right]=\delta_{\alpha\beta}$.}
	\label{tab:multi}
\end{table}

\noindent {\color{blue} \it Introduction}--
Weyl fermions, originally proposed in high-energy physics~\cite{Weyl1929}, are recently discovered in solid-state materials as linearly dispersed low-energy quasiparticles~\cite{Hasan2017,Yan2017,Burkov2018,Armitage2018,Ding2021}. This discovery has aroused a flurry of research interests for their nontrivial band topology.
In reciprocal space, Weyl nodes characterized by the chirality $\pm1$ correspond to the source ($+$) or sink ($-$) of Berry curvature flux~\cite{Berry1984,Bohm2003,Niu2010}, which renders a rich variety of unusual phenomena covering the chiral anomaly~\cite{Adler1969,Bell1969,Nielsen1983,Zyuzin2012,Hosur2012,Kim2013,Son2013,Parameswaran2014,Burkov2014a,Chernodub2014,Burkov2015,Huang2015a,Hirschberger2016,Zhang2016,Arnold2016,Wang2016b,Xie2016,Niemann2017}, anomalous Hall effect~\cite{Yang2011,Burkov2014b,Steiner2017,Liang2018,Wang2018,Liu2018,Jiang2021}, and Fermi-arc surface states~\cite{Wan2011,Xu2011,Xu2018,Xie2017,Xie2020}. From the symmetry perspective, one promising realization of Weyl fermions is to lift the Kramers degeneracy of parent systems by breaking either the spatial inversion~\cite{Dai2015,Huang2015b,Lv2015a,Lv2015b,Xu2015a,Xu2015b,Xu2015c,Xu2016,Xu2017,Lai2017,Dzsaber2021} or time reversal symmetry~\cite{Wang2016a,Suzuki2016,Yang2017,Jin2017,Shekhar2018,Liu2019,Morali2019,Borisenko2019,Nie2020}. In particular, the spin degeneracy of electronic bands for the latter case is removed as the system undergoes a magnetic transition, beyond which Weyl fermions emerge. It represents a rare example that the intersection of band topology and electronic correlation breeds emergent topological phases.

Here, we present a complementary approach to realize Weyl fermions in a $p$-band diamond lattice with the driving force arsing from the correlation among orbital multiplets. Unlike the $\text{SU}(2)$ symmetry of spins, the orbital anisotropy characterized by the spatial orientation breaks the orbital rotational symmetry through the coupling of orbitals to the host lattice geometry. Hence, the orbital multipolar ordering of an interacting system ultimately determines the underlying space group symmetry of host lattice, thereby dictating the band degeneracy in reciprocal space. Interestingly, we show that Weyl fermions with nontrivial band topology can emerge from an orbital multipolar ordering phase through a detailed symmetry analysis on the evolution of degenerate band nodes. Our study also provides a representative recipe that the diverse orbital multipolar orderings enrich the realization of emergent novel fermions in correlated multiorbital systems.

\begin{figure}
	\centering
	\includegraphics[width=0.5\textwidth]{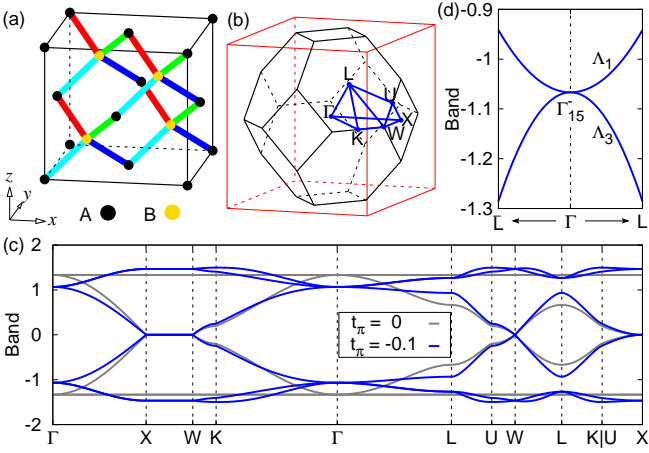}	
	\caption{ 
		(a) The bipartite structure of diamond lattice and (b) the first Brillouin zone with Wigner-Seitz constructions. 
		(c) The band structure of tight-binding model in Eq.~(\ref{eq:TB}) along the high symmetry lines indicated in (b) with $t_\sigma=1$.
		(d) The lower three bands around $\Gamma$ point with $t_\pi=-0.1$. The triply degenerate band node with representation $\Gamma_{15}^-$ of group $O_h$ is split into a one dimensional representation $\Lambda_1$ and a two dimensional representation $\Lambda_3$ of group $C_{3v}$ along high symmetry line $\bar{\text{L}}-\Gamma-\text{L}$.
	}
	\label{fig:band}
\end{figure}

\noindent {\color{blue} \it Minimal Tight-Binding Model}--
We start with the tight-binding model that describes the hopping process in the $p$-band diamond lattice. This model is previously proposed in the study of frustrated orbital superfluids~\cite{Chern2014}. As depicted in Fig.~\ref{fig:band}(a), the diamond lattice contains two superimposed face-centered-cubic sublattices A and B with displacement $\bm{e}_4=a/2\left(-1,-1,-1\right)$. Its symmetry is characterized by the nonsymmorphic space group $O_h^7$ with sublattice interchange $\{\mathcal{I}|\bm{e}_4\}$ involving the inversion $\mathcal{I}$ followed by a fractional translation of primitive lattice vectors $\bm{e}_4$. Introducing an orbital-sublattice field operator, the tight-binding Hamiltonian in reciprocal space reads
\begin{eqnarray}
	H_{\text{TB}} &&= \sum_{\bm k} \sum_{\mu,\nu=x,y,z} T^{\mu\nu}_{\bm k} 
	p^\dagger_{\mu \text{A} \bm k} p_{\nu \text{B}\bm k}+\text{h.c.},
	\label{eq:TB}	\\
	&&T^{\mu\nu}_{\bm k} = \left[t_\sigma+\left(3\delta_{\mu\nu}-1\right)t_\pi\right]\sum_{i}\epsilon^{\mu\nu}_{i\bm{k}},
	\nonumber
\end{eqnarray}
where the dispersion $\epsilon^{\mu\nu}_{i\bm{k}} = \hat{e}^\mu_i \hat{e}^\nu_i \exp\left[-i\bm{k}\left(\bm{e}_i-\bm{e}_4\right)\right]$ describes the quantum tunnelling from $p_{\nu}$ to $p_{\mu}$ orbitals along the $i$-th bond vector $\bm{e}_i$. Here we have chosen the sublattices A and B connected by the bond vector $\bm{e}_4$ as a primitive unit cell. The hopping integrals $t_\sigma$ and $t_\pi$ denote the $\sigma$ and $\pi$ bondings of $p$ orbitals, respectively. For the $\pi$ bonding, the bond vector lies in the nodal plane of $p$ orbitals. Consequently, the strength of $\pi$ bonding is much weaker than that of $\sigma$ bonding. The band structure of the tight-binding model in Eq.~(\ref{eq:TB}) with $t_\sigma=1$, plotted in Fig.~\ref{fig:band}(c), is symmetric with respect to zero energy across the first Brillouin zone (FBZ), originating from the chiral symmetry $\Xi=\tau_z$. Here $\tau_z$ is the $z$-component Pauli matrix operating on the sublattice degree of freedom. Notably, the lower and upper three bands at $\Gamma$ point are triply degenerate, pinning the Fermi level at the commensurate filling of $\nu=1$ and $2$ per site, respectively. These degeneracies root in symmetry. The reciprocal space  representations at $\Gamma$ point, having the highest symmetry in FBZ, can be decomposed into the irreducible representations $\Gamma_1^+$ and $\Gamma_2^-$ of cubic space group $O_h$ with the superscript $\pm$ referring to even or odd parity of inversion~\cite{Dresselhaus2008}. Taking the $p$-orbital representation $\Gamma_{15}^-$ into account, the lower and upper three bands are thus described by irreducible representations $\left(\Gamma_1^+\oplus\Gamma_2^-\right)\otimes\Gamma_{15}^-=\Gamma_{15}^-\oplus\Gamma_{25}^{+}$, which correspond to the anti-bonding and bonding of $p$ orbitals, respectively. Hereafter we shall only focus on the lower bands since the physics of upper bands can be easily derived through the chiral operation $\Xi$. To describe the low-energy behavior, we next construct an effective three-band $k\cdot p$ model around $\Gamma$ point. Under cubic space group $O_h$, the wave vector $\bm{k}$ transforms as a polar vector with representation $\Gamma_{15}^-$. In contrast, the density matrix spanned by the anti-bonding $p$ orbitals can be decomposed in terms of orbital angular momenta $\hat{\bm{l}}$~\cite{Blum2012}, which transform as an axial vector with representation $\Gamma_{15}^+$. As tabulated in Table~\ref{tab:multi}, the detailed decomposition is classified as monopole (charge) $\hat{n}$, dipole $\bm{\hat{J}}$, and quadrupole $\bm{\hat{O}}$, which transform under rotations as effective angular momenta $\ell=0,1,2$ respectively~\cite{Santini2009}. The effective three-band $k\cdot p$ model is given by 
\begin{eqnarray}
	\mathcal{H}_{\Gamma} \left(\bm{k}\right) &=& -\frac{4}{3}\left(t_\sigma+2t_\pi\right)
	+d_{\Gamma_1^+}\left(\bm{k}\right)\hat{n} \nonumber\\
	&+&\bm{d}_{\Gamma_{25}^+}\left(\bm{k}\right)\cdot\hat{\bm{O}}_{\Gamma_{25}^+}
	+\bm{d}_{\Gamma_{12}^+}\left(\bm{k}\right)\cdot\hat{\bm{O}}_{\Gamma_{12}^+}+\mathcal{O}\left(k^4\right)
	\label{eq:kp}
\end{eqnarray}
where the operators $\hat{\bm{O}}_{\Gamma_{25}^+}=\left(\hat{O}_{yz},\hat{O}_{zx},\hat{O}_{xy}\right)$ 
and $\hat{\bm{O}}_{\Gamma_{12}^+}=\left(\hat{O}_{x^2-y^2},\hat{O}_{3z^2-r^2}\right)$ are the symmetry classified quadrupole moments in $\Gamma_{25}^+$ and $\Gamma_{15}^+$ representations respectively, and the coefficients
\begin{eqnarray}
	d_{\Gamma_1^+}\left(\bm{k}\right) &=& \left[\frac{t_\sigma+2t_\pi}{2}
	-\frac{1}{3}\frac{\left(t_\sigma-t_\pi\right)^2}{\left(t_\sigma+2t_\pi\right)}\right]\frac{k^2}{\sqrt{3}},	\nonumber\\
	\bm{d}_{\Gamma_{25}^+}\left(\bm{k}\right) &=& -\frac{\sqrt{2}}{6}\frac{\left(t_\sigma-t_\pi\right)\left(t_\sigma+5t_\pi\right)}{t_\sigma+2t_\pi}
	\left(k_yk_z,k_zk_x,k_xk_y\right),	\nonumber\\
	\bm{d}_{\Gamma_{12}^+}\left(\bm{k}\right) &=& -\frac{\sqrt{2}}{6}\frac{\left(t_\sigma-t_\pi\right)^2}{t_\sigma+2t_\pi}
	\left(\frac{3k_z^2-k^2}{\sqrt{6}},\frac{k_x^2-k_y^2}{\sqrt{2}}\right).	\nonumber
\end{eqnarray}
See Supplemental Material~\cite{SM} for details of derivations. Under the symmetry transformation, the effective Hamiltonian in Eq.~(\ref{eq:kp}) is kept invariant through a dual irreducible representation of cubic group symmetry $O_h$ between the coefficients $\bm{d}\left(\bm{k}\right)$ and multipole operators $\hat{\bm{O}}$ from symmetry aspects~\cite{Lew2009}. It ensures that the linear order in $\bm{k}$ and $\hat{\bm{l}}$ vanishes identically due to the opposite parties. Therefore, the quadratic order that transforms as  $\Gamma_{15}^\pm\otimes\Gamma_{15}^\pm=\Gamma_1^+\oplus\Gamma_{12}^+\oplus\Gamma_{15}^+\oplus\Gamma_{25}^+$ emerges as the leading order. Notably, the effective Hamiltonian only involves time-reversal-invariant multipole operators, which ensures that the Berry curvature is odd in reciprocal space. Moreover, it also preserves the inversion symmetry $\mathcal{H}_\Gamma\left(\bm{k}\right)=\mathcal{H}_\Gamma\left(-\bm{k}\right)$ originating from the aforementioned sublattice symmetry. Hence, the Berry curvature vanishes strictly due to these two symmetries. As we will show later, the spontaneous breaking of time reversal symmetry induced by many-particle interactions renders a nontrivial band topology. Generally, diagonalizing $\mathcal{H}_\Gamma\left(\bm{k}\right)$ yields a triply degenerate quadratic band node for $t_\pi\ne0$. Concise results can be analytically obtained along the high symmetry line $\bar{\text{L}}-\Gamma-\text{L}$. The reciprocal space group is reduced to group $C_{3v}$ with rotation axis along $\left[111\right]$ direction. Accordingly, as shown in Fig.~\ref{fig:band}(d) the quadratic band node is split into a band
$E_{\Lambda_1}\left(\bm{k}\right)=-4\left(t_\sigma+2t_\pi\right)/3+t_\sigma\left(t_\sigma+8t_\pi\right)k^2/6\left(t_\sigma+2t_\pi\right)$ in representation $\Lambda_1$, and doubly degenerate bands  $E_{\Lambda_3}\left(\bm{k}\right)=-4\left(t_\sigma+2t_\pi\right)/3+t_\pi\left(4t_\sigma+5t_\pi\right)k^2/6\left(t_\sigma+2t_\pi\right)$ in representation $\Lambda_3$. In the limit $t_\pi=0$, lowest two bands are completely flat [c.f. grey lines in Fig.~\ref{fig:band}(c)], which leads to the itinerant ferromagnetism for spin-$1/2$ fermions as studied below. In $p$-band honeycomb lattices, the flat band with a highly degenerate manifold of single-particle states promotes various emergent many-particle states by invoking interactions~\cite{Wu2007,Chen2019}.  

\begin{figure}
	\centering
	\includegraphics[width=0.5\textwidth]{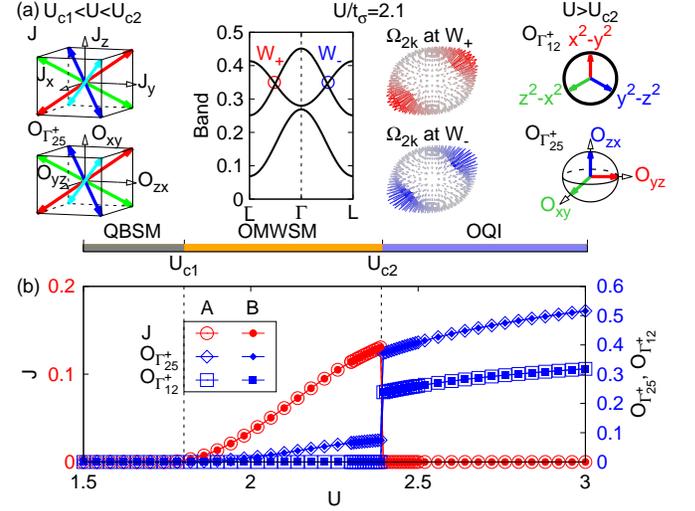}	
	\caption{ 
	(a) The phase diagram as a function of Hubbard interaction $U$ at $\{t_\sigma,t_\pi\}=\{1,-0.1\}$ shows three phases:
	(1) quadratic band semimetal (QBSM),
	(2) orbital multipolar Weyl semimetal (OMWSM) with the dipolar $\bm{J}$ and quadrupolar $\bm{O}_{\Gamma_{25}^+}$ orderings aligned along one of the bond vectors,
	(3) orbital quadrupolar insulator (OQI) with the quadrupolar orderings $\bm{O}_{\Gamma_{25}^+}$ and $\bm{O}_{\Gamma_{15}^+}$.
    The preferable orientation of orbital multipole due to spontaneous symmetry breaking are indicated. 
    The Berry curvature around the Weyl nodes is shown to characterize the topology of OMWSM.  
	(b) The evolution of orbital multipole magnitudes. 
	There are two boundaries, $U_{c1}$ and $U_{c2}$, separating QBSM, OMWSM, and OQI, respectively.
	All multipolar orderings show ferro-orbital correlations, except the case that $O_{\Gamma_{25}^+}$ has an antiferro-orbital correlation beyond the critical interaction $U_{c2}$. 
	}
	\label{fig:hf}
\end{figure}

\noindent {\color{blue} \it Interacting Spinless Fermions}--
The quadratic band node in three dimensions has been shown as a fertile ground to host various interacting topological phases~\cite{Kondo2015}. In the following, we will show that a topological semimetal with Weyl fermions can emerge purely from the correlations of orbital multiplets by studying the interacting spinless fermions. According to the Fermi statistics, the interaction between $p$-orbital fermions merely arises from $p$-wave channel and takes the usual form of Hubbard interaction~\cite{Wu2011}. The interacting Hamiltonian, at mean-field level~\cite{Flensberg2004}, is described by Hartree and multipole exchange self-energies
\begin{equation}
	H_\text{I}
	=U\sum_i\left(2n_i\hat{n}_i
	-\bm{J}_i\cdot\hat{\bm{J}}_i
	-\bm{O}_i\cdot\hat{\bm{O}}_i
	-\frac{2n_i^2-\bm{J}_i^2-\bm{O}_i^2}{2}\right) 
	\label{eq:HF}
\end{equation}
where $n_i$, $\bm{J}_i$ and $\bm{O}_i$ are the ground-state expectation values of multipole operators at the $i$-th site. The first term in Eq.~(\ref{eq:HF}) renormalizes the on-site energy level, while the dipole and quadrupole exchange interactions, the second and third terms, preserve the orbital rotation symmetry and favor the multipolar order by lowering the exchange self-energy. The ground state is obtained by self-consistently solving the mean-field Hamiltonian composed of the tight-binding part in Eq.~(\ref{eq:TB}) and the interacting part in Eq.~(\ref{eq:HF}). As sketched in Fig.~\ref{fig:hf}(a), the calculated phase diagram with the hopping integrals $\{t_\sigma,t_\pi\}=\{1,-0.1\}$ accommodates three different phases including quadratic band semimetal (QBSM), orbital multipolar Weyl semimetal (OMWSM), and orbital quadrupolar insulator (OQI). Figure~\ref{fig:hf}(b) plots the evolution of orbital multipole magnitude for both A and B sublattices. Across the whole phase diagram, the sublattices A and B develop an identical magnitude, preserving the sublattice symmetry. Initially, a weak Hubbard interaction beyond $U_{c1}$ drives an orbital multipolar order intertwining both the dipole $\bm{J}$ and quadrupole $\bm{O}_{\Gamma^+_{25}}$ moments aligned one of bond vectors. The quadrupole is time-reversal invariant, while the dipole is time-reversal odd. Therefore, this phase spontaneously breaks the time-reversal symmetry, which renders nontrivial band topology. Specifically, the dipole $\bm{J}$ with a spontaneously selected direction, {\it e.g.} [111] axis, breaks the aforementioned twofold band degeneracy in representation $\Lambda_3$ of group $C_{3v}$ along the high symmetry line $\bar{\text{L}}-\Gamma-\text{L}$, accompanied by two Weyl nodes crossed with the high-lying band in representation $\Lambda_1$. The nontrivial band topology of OMWSM phase is characterized by the Berry curvature around the Weyl nodes
\begin{equation}
	\bm{\Omega}_{n\bm{k}}=-\text{Im}\sum_{n^\prime\ne n}
	\frac{\left<n\bm{k}\left|\nabla_{\bm{k}}\mathcal{H}_{\bm{k}}\right|n^\prime\bm{k}\right>
	\times\left<n^\prime\bm{k}\left|\nabla_{\bm{k}}\mathcal{H}_{\bm{k}}\right|n\bm{k}\right>}
	{\left(E_{n\bm{k}}-E_{n^\prime\bm{k}}\right)^2} \nonumber
\end{equation}
where $E_{n\bm{k}}$ and $\left|n\bm{k}\right>$ are the $n$-th eigenvalue and eigenvector of Bloch Hamiltonian $\mathcal{H}_{\bm{k}}$, respectively.
The calculated Berry curvature of second band $\bm{\Omega}_{2\bm{k}}$ shown in Fig.~\ref{fig:hf}(a) is strongly anisotropy with large weight distributed along the spontaneously selected direction of multipole $\bm{J}$ and $\bm{O}_{\Gamma_{25}^+}$ moments. The chirality of Weyl nodes is characterized by the Chern number, which can be calculated by integrating the Berry curvature of $n$-th band 
\begin{eqnarray}
	\text{Ch}_n=\frac{1}{2\pi}\oint_{\text{FS}}\bm{\Omega}_{n\bm{k}}\cdot d\bm{k}
\end{eqnarray}
over a closed Fermi surface that encloses the Weyl nodes. The evaluated Chern numbers suggest that these two Weyl nodes have different chirality, indicating that each of Weyl nodes carries a $\pm2\pi$ Berry flux. Since Weyl fermions must come in pairs~\cite{Nielsen1981}, this phase serves as a minimal setting with the number of Weyl nodes being two. With further raising the Hubbard interaction $U$, the system undergoes a phase transition from OMWSM into OQI, which is expected to be first-order type due to distinct broken symmetries of these two phases. Hence, the corresponding phase boundary is indicated by the discontinuous jump of multipole magnitude. In this phase, the correlation of quadrupolar order $\bm{O}_{\Gamma_{25}^+}$ switches to an antiferro-orbital order between A and B sublattices, accompanied by a ferro-orbital quadrupolar order $\bm{O}_{\Gamma_{12}^+}$.

\begin{figure}
	\centering
	\includegraphics[width=0.5\textwidth]{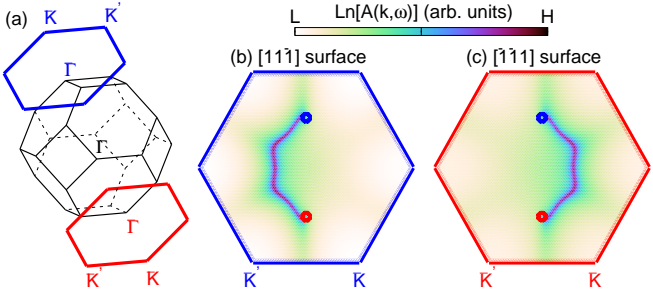}	
	\caption{
		(a) Schematic plot of projected [$11\bar{1}$] and [$\bar{1}\bar{1}1$] surfaces Brillouin zone.
		The calculated surface spectral weight $A\left(\bm{k},\omega\right)$ in Eq.~(\ref{eq:spectral}) at Fermi energy shows that Fermi arcs connect a pair of projected Weyl nodes with opposite chirality on (b) [$11\bar{1}$] and (c) [$\bar{1}\bar{1}1$] surfaces. The red and blue dots correspond to the surface projection of Weyl nodes with chirality $+1$ and $-1$, respectively.	The color bar indicates the high (H) and low (L) spectral weight distributions on the surfaces. 
	}
	\label{fig:fa}
\end{figure}

The topological nature of OMWSM phase can be further demonstrated by topologically protected Fermi arcs through the bulk-boundary correspondence~\cite{Wan2011,Xu2011}. To this end, the surface spectral function of a semi-infinite system is calculated through the retarded surface
Green function technique~\cite{Sancho1984,Sancho1985}
\begin{equation}
	A\left(\bm{k},\omega\right) = -\frac{1}{\pi}\text{Im}
	\left[\text{Tr}\text{G}^{\text{R}}\left(\bm{k},\omega\right)\right].
	\label{eq:spectral}
\end{equation}
Figures~\ref{fig:fa}(b) and \ref{fig:fa}(c) show the numerical evaluation of spectral weight on the [$11\bar{1}$] and [$\bar{1}\bar{1}1$] surfaces at Fermi energy, respectively. The open Fermi arcs connecting the projections of Weyl nodes with opposite chirality shift oppositely along the transverse direction on the top and bottom surfaces in Fig.~\ref{fig:fa}(a).

\begin{figure}
	\centering
	\includegraphics[width=0.5\textwidth]{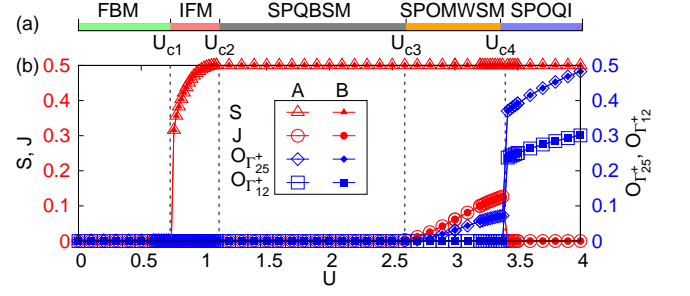}	
	\caption{ 
		(a) The phase diagram as a function of intra-orbital Coulomb interaction $U$ at $\{t_\sigma,t_\pi,J_H/U\}=\{1,-0.1,0.1\}$ shows five phases:
		(1) flat-band metal (FBM),
		(2) itinerant ferromagnetic phase (IFM),
		(3) spin-polarized quadratic band semimetal (SPQBSM),
		(4) spin-polarized orbital multipolar Weyl semimetal (SPOMWSM) with the dipolar $\bm{J}$ and quadrupolar $\bm{O}_{\Gamma_{25}^+}$ orderings,
		(5) spin-polarized orbital quadrupolar insulator (SPOQI) with the quadrupolar orderings $\bm{O}_{\Gamma_{25}^+}$ and $\bm{O}_{\Gamma_{15}^+}$.
		(b) The evolution of spin and orbital multipole magnitudes, which are defined with summation over the orbital and spin indices respectively. 
	}
	\label{fig:shf}
\end{figure}

\noindent {\color{blue} \it Interacting Spin-$1/2$ Fermions}--
Having established the phase diagram of interacting spinless fermions, we are then in a position to generalize our study to spin-$1/2$ fermions, which are relevant for solid-state materials. The multiorbital correlations are described by 
\begin{eqnarray}
	H_\text{I} &=& U\sum_{i,\mu}\hat{n}_{i\mu\uparrow}\hat{n}_{i\mu\downarrow}
               + \left(U^\prime-\frac{1}{2}J_H\right)\sum_{i,\mu<\nu}\hat{n}_{i\mu}\hat{n}_{i\nu} \nonumber\\
               &-& J_H\sum_{i,\mu\ne\nu}\bm{S}_{i\mu}\cdot\bm{S}_{i\nu}
               + J_H\sum_{i,\mu\ne\nu}p^\dagger_{i\mu\uparrow}p^\dagger_{i\mu\downarrow}
               p_{i\nu\downarrow}p_{i\nu\uparrow}                         	
\end{eqnarray}
where $\hat{n}_{i\mu}$ and $\bm{S}_{i\mu}$ are charge and spin operators in orbital $p_{\mu}$ at $i$-th site respectively, and the intra- and inter-orbital Coulomb interactions $\left(U,U^\prime\right)=\left(F^0+4F^2/25,F^0/-2F^2/25\right)$ are related to Hund's coupling $J_H$ by $U^\prime=U-2J_H$. The orbital rotation symmetry is fully respected with the Slater-Condon-Shortley parameterization $F^0$ and $F^2$~\cite{Slater1929,Condon1931}. See Supplemental Material~\cite{SM} for details of derivations. The calculated phase diagram within the standard mean-field approximation is shown in Figs.~\ref{fig:shf}(a) and \ref{fig:shf}(b). Initially, the system is in a metallic phase with half-filled flat bands. Tiny interactions drive the system into itinerant ferromagnetic (IFM) phase at critical interaction $U_{c1}$, which is attributed to the energy reduction of repulsive interactions. The flat-band induced IFM phase is also found in a $p$-band honeycomb lattice by the early study~\cite{Zhang2010}. Recently, the flat-band ferromagnetism is experimentally observed in a multi $d$-orbital system~\cite{Lin2018,Ye2018,Yin2018}. With further raising $U$, the spin is fully polarized at critical interaction $U_{c2}$, leaving that the orbital degree of freedom is only activated. Therefore, the ensuing transitions into OMWSM and OQI phases but with the spin fully polarized resembles those of spinless fermions, demonstrating that these transitions are orbital driven in nature.

\noindent {\color{blue} \it Concluding Remarks}--
To summarise, we have predicted a Weyl semimetal phase emergent from the orbital multipolar orderings for both spinless and spin-$1/2$ fermions. An advantage of our study is that the orbital anisotropy promoted by the orbital multipolar orderings is locked to the host lattice geometry, which may stabilize the predicted Weyl semimetals in return. 
Moreover, high-rank multipoles, {\it e.g.} octupoles in $d$-orbital systems, are expected to enrich the evolution of multifold degenerate band nodes with more complexity, which remains open for future study. Our study illustrates the essential ingredients of design principles to guide the search of emergent novel fermions with nontrivial band topology in correlated multiorbital systems.

\section{Acknowlegdgements}
This work is supported by NSFC under Grants No. 12174345, 12174317, 11729402, 11704338, 11534001, and NBRPC under Grant No. 2015CB921102. 


%


\widetext

\setcounter{equation}{0}
\setcounter{figure}{0}
\setcounter{table}{0}
\renewcommand{\theequation}{S\arabic{equation}}
\renewcommand{\thefigure}{S\arabic{figure}}
\renewcommand{\thetable}{S\arabic{table}}
\renewcommand{\thesubsection}{S\arabic{subsection}}


\section{SUPPLEMENTAL MATERIALS}
We present the detailed information about 
(${\rm\uppercase\expandafter{\romannumeral1}}$) The second-order effective Hamiltonian,
and (${\rm\uppercase\expandafter{\romannumeral2}}$) The multiorbital Hubbard interaction.

\subsection{I. The second-order effective Hamiltonian}

Introducing an orbital-sublattice spinor representation $\bm{p}_{\bm k}=\left[p_{x\text{A}{\bm k}},p_{y\text{A}{\bm k}},p_{z\text{A}{\bm k}},p_{x\text{B}{\bm k}},p_{y\text{B}{\bm k}},p_{z\text{B}{\bm k}},\right]^\text{T}$, 
the tight-binding model that describes the hopping process in the $p$-band diamond lattice has the following form
\begin{equation}
	H_{\text{TB}} = \sum_{\bm k} \bm{p}^\dagger_{\bm k} \mathscr{H}_{\bm k} \bm{p}_{\bm k},
	\mathscr{H}_{\bf k}=\left[
	\begin{matrix}
		0 & T_{\bm k} \\
		T^\dagger_{\bm k} & 0
	\end{matrix}
	\right]. 
	\label{eq:TB-S}
\end{equation}
Here the elements of matrix $T_{\bm{k}}$ are given by 
\begin{equation}
	T^{\mu\nu}_{\bm k} = \left[t_\sigma+\left(3\delta_{\mu\nu}-1\right)t_\pi\right]\sum_{i}\epsilon^{\mu\nu}_{i\bm{k}}. 
\end{equation}
with the dispersion $\epsilon^{\mu\nu}_{i\bm{k}} = \hat{e}^\mu_i \hat{e}^\nu_i \exp\left[-i\bm{k}\left(\bm{e}_i-\bm{e}_4\right)\right]$ describing the quantum tunnelling from $p_{\nu}$ to $p_{\mu}$ orbitals along the $i$-th bond vector $\bm{e}_i$. In Eq.~(\ref{eq:TB-S}), the chiral symmetry of the Bloch Hamiltonian becomes transparent
\begin{eqnarray}
	\Xi\mathscr{H}_{\bm{k}}\Xi=-\mathscr{H}_{\bm{k}},
	\label{eq:Chiral-S}
\end{eqnarray}
 where the chiral operator $\Xi=\tau_z$ is the $z$-component Pauli matrix operating on the sublattice degree of freedom. If $|\psi_{\bm{k}}^n\rangle$ is an eigenstate of the Bloch Hamiltonian $\mathcal{H}_{\bm{k}}$ with energy $E_{\bm{k}}^n$, there is a partner eigenstate $\Xi|\psi_{\bm{k}}^n\rangle$ satisfying 
 \begin{eqnarray}
 	\mathscr{H}_{\bm{k}}\Xi|\psi_{\bm{k}}\rangle
 	=-\Xi\mathscr{H}_{\bm{k}}|\psi_{\bm{k}}\rangle
 	=-E_{\bm{k}}^n|\psi_{\bm{k}}\rangle
 \end{eqnarray} 
Therefore, the band structure of Bloch Hamiltonian is symmetric with respect to zero energy. 

At the Brillouin zone center $\Gamma$ point, the band dispersion has two sets of threefold degeneracy with eigen energies $E_\Gamma^\pm=\pm\frac{4}{3}\left(t_\sigma+2t_\pi\right)$, 
which pin exactly the Fermi levels at filling $\nu=2$ and $\nu=1$ per site, respectively.
The eigen vectors of the upper $E_\Gamma^+$ and lower $E_\Gamma^-$ eigen energies are given by the bonding ($\psi_\Gamma^{+\mu}$) and anti-bonding ($\psi_\Gamma^{-\mu}$) states of $p$ orbitals
\begin{eqnarray}
	\psi_\Gamma^{\pm\mu} &=& \frac{1}{\sqrt{2}}\left(p_{\mu\text{A}\Gamma}\pm p_{\mu\text{B}\Gamma}\right).
\end{eqnarray}	
In this basis, the Hamiltonian expanded around $\Gamma$ point reads
\begin{eqnarray}
	\mathcal{H}_{\bm{k}}
	=\left[
	\begin{matrix}
		E_{\Gamma}^- -T_{\bm{k}}^+ & T_{\bm{k}}^- \\
		-T_{\bm{k}}^- & E_{\Gamma}^+ + T_{\bm{k}}^+
	\end{matrix}
	\right]	
	\label{eq:HG-S}
\end{eqnarray}
where the auxiliary matrices
\begin{eqnarray}
	T_{\bm{k}}^\pm=\frac{1}{2}\left(T_{\bm{k}} \pm T_{\bm{k}}^\dagger\right).
\end{eqnarray}
The upper ($E_\Gamma^+$) and lower ($E_\Gamma^-$) bands at $\Gamma$ point are well separated in energy by a band gap $\Delta_{\Gamma}=\frac{8}{3}\left(t_\sigma+2t_\pi\right)$.
The low-energy behavior around $\Gamma$ point at filling $\nu=1$ ($2$) is renormalized by a second-order virtual process in which the fermion first hops from the lower (upper) bands to the upper (lower) bands and then hops back to the lower (upper) bands. Mathematically, the effective Hamiltonian can be obtained by eliminating the off-diagonal elements of Eq.~(\ref{eq:HG-S}) through a canonical transformation 
\begin{eqnarray}
	\mathcal{H}_{\Gamma}\left(\bm{k}\right)=\exp\left[-Y_{\bm{k}}\right]\mathcal{H}_{\bm{k}}\exp\left[Y_{\bm{k}}\right],
	Y_{\bm{k}}=\left[
	\begin{matrix}
		0 & X_{\bm{k}} \\
		-X_{\bm{k}}^\dagger & 0
	\end{matrix}
	\right]	
\end{eqnarray}
where the matrix $X_{\bm{k}}$ is determined by
\begin{eqnarray}
	T_{\bm{k}}^-+E_{\Gamma}^-X_{\bm{k}}-X_{\bm{k}}E_{\Gamma}^+=0.
\end{eqnarray}
Having uniquely determined $X_{\bm{k}}$, the derivation of the effective Hamiltonian is now straightforward
\begin{eqnarray}
	\mathcal{H}_{\Gamma}\left(\bm{k}\right)
		=\left[
	\begin{matrix}
		\mathcal{H}_{\Gamma}^{-}\left(\bm{k}\right)  & 0 \\
		0 & \mathcal{H}_{\Gamma}^{+}\left(\bm{k}\right) 
	\end{matrix}
	\right]	,
\end{eqnarray}
where
\begin{eqnarray}
	\mathcal{H}_{\Gamma}^{\pm}\left(\bm{k}\right) 
	= E_{\Gamma}^{\pm} \pm T_{\bm{k}}^+ \mp \frac{1}{\Delta_{\Gamma}}T_{\bm{k}}^- T_{\bm{k}}^-
	\label{eq:HPM-S}
\end{eqnarray}
describes the low-energy behavior of lower ($-$) and upper ($+$) three bands around $\Gamma$ point. This result recovers the well-known Brillouin-Wigner perturbation theory~\cite{Hubac2010}.
Finally, the expression of effective Hamiltonian in the main text can be easily derived by explicitly expanding $\bm{T}_{\bm{k}}^\pm$ in powers of $\bm{k}$ after a lengthy but straightforward algebraic calculation.
It is worthy mentioning that the effective three-band $k\cdot p$ Hamiltonian $\mathcal{H}_{\Gamma}^{\pm}\left(\bm{k}\right)$ in Eq.~(\ref{eq:HPM-S}) can be transformed to each other through the chiral operation $\Xi$.

\begin{table*}
	\begin{ruledtabular}
		\begin{tabular}{c|ccccc}
			No.	&$\mu_1,\sigma_1$	&$\mu_2,\sigma_2$	&$\mu_3,\sigma_3$	&$\mu_4,\sigma_4$	&$I(\mu_1\sigma_1,\mu_2\sigma_2;\mu_3\sigma_3,\mu_4\sigma_4)$	\\ 
			\hline
			1	& 1 , -1 & 2 , -1 & 1 , -1 & 2 , -1 & -$F^0$ + $F^2/5$ \\ 
			2	& 1 , -1 & 3 , -1 & 1 , -1 & 3 , -1 & -$F^0$ + $F^2/5$ \\ 
			3	& 1 , -1 & 1 , 1 & 1 , -1 & 1 , 1 & -$F^0$ - $4F^2/25$ \\ 
			4	& 1 , -1 & 1 , 1 & 2 , -1 & 2 , 1 & -$3F^2/25$ \\ 
			5	& 1 , -1 & 1 , 1 & 3 , -1 & 3 , 1 & -$3F^2/25$ \\ 
			6	& 1 , -1 & 2 , 1 & 1 , -1 & 2 , 1 & -$F^0$ + $2F^2/25$ \\ 
			7	& 1 , -1 & 2 , 1 & 2 , -1 & 1 , 1 & -$3F^2/25$ \\ 
			8	& 1 , -1 & 3 , 1 & 1 , -1 & 3 , 1 & -$F^0$ + $2F^2/25$ \\ 
			9	& 1 , -1 & 3 , 1 & 3 , -1 & 1 , 1 & -$3F^2/25$ \\ 
			10	& 2 , -1 & 3 , -1 & 2 , -1 & 3 , -1 & -$F^0$ + $F^2/5$ \\ 
			11	& 2 , -1 & 1 , 1 & 1 , -1 & 2 , 1 & -$3F^2/25$ \\ 
			12	& 2 , -1 & 1 , 1 & 2 , -1 & 1 , 1 & -$F^0$ + $2F^2/25$ \\ 
			13	& 2 , -1 & 2 , 1 & 1 , -1 & 1 , 1 & -$3F^2/25$ \\ 
			14	& 2 , -1 & 2 , 1 & 2 , -1 & 2 , 1 & -$F^0$ - $4F^2/25$ \\ 
			15	& 2 , -1 & 2 , 1 & 3 , -1 & 3 , 1 & -$3F^2/25$ \\ 
			16	& 2 , -1 & 3 , 1 & 2 , -1 & 3 , 1 & -$F^0$ + $2F^2/25$ \\ 
			17	& 2 , -1 & 3 , 1 & 3 , -1 & 2 , 1 & -$3F^2/25$ \\ 
			18	& 3 , -1 & 1 , 1 & 1 , -1 & 3 , 1 & -$3F^2/25$ \\ 
			19	& 3 , -1 & 1 , 1 & 3 , -1 & 1 , 1 & -$F^0$ + $2F^2/25$ \\ 
			20	& 3 , -1 & 2 , 1 & 2 , -1 & 3 , 1 & -$3F^2/25$ \\ 
			21	& 3 , -1 & 2 , 1 & 3 , -1 & 2 , 1 & -$F^0$ + $2F^2/25$ \\ 
			22	& 3 , -1 & 3 , 1 & 1 , -1 & 1 , 1 & -$3F^2/25$ \\ 
			23	& 3 , -1 & 3 , 1 & 2 , -1 & 2 , 1 & -$3F^2/25$ \\ 
			24	& 3 , -1 & 3 , 1 & 3 , -1 & 3 , 1 & -$F^0$ - $4F^2/25$ \\ 
			25	& 1 , 1 & 2 , 1 & 1 , 1 & 2 , 1 & -$F^0$ + $F^2/5$ \\ 
			26	& 1 , 1 & 3 , 1 & 1 , 1 & 3 , 1 & -$F^0$ + $F^2/5$ \\ 
			27	& 2 , 1 & 3 , 1 & 2 , 1 & 3 , 1 & -$F^0$ + $F^2/5$ \\ 
		\end{tabular}
	\end{ruledtabular}
	\caption{
		Matrix elements of the Slater-Condon-Shortley parameterized Coulomb interaction. 
		Repeated elements derived from the relation 
		$ I\left(\mu_1\sigma_1,\mu_2\sigma_2;\mu_3\sigma_3,\mu_4\sigma_4\right)
		= I\left(\mu_2\sigma_2,\mu_1\sigma_1;\mu_4\sigma_4,\mu_3\sigma_3\right)
		=-I\left(\mu_1\sigma_1,\mu_2\sigma_2;\mu_4\sigma_4,\mu_3\sigma_3\right)
		=-I\left(\mu_2\sigma_2,\mu_1\sigma_1;\mu_3\sigma_3,\mu_4\sigma_4\right)$
		are not listed.
		In the table, the spin indices $\sigma=1$ and $-1$ label the $\uparrow$ and $\downarrow$ spin states,
		and the orbital indices $\mu=1,2,3$ label the $p_x,p_y,p_z$ orbital states, respectively. 
	}
	\label{tab:Slater-S}
\end{table*}

\subsection{II. The multiorbital Hubbard interaction}
In this section, we will formulate the many-particle Hamiltonian to describe the repulsive Coulomb interaction for $p$ orbital electrons. The derivation is well established in the study of $d$-orbital electronic materials~\cite{Georges2013}. For an isolated atom with open $nl$ shell, the electronic wave function takes the following form
\begin{eqnarray}
	\phi_{m\sigma}\left(\bm{r}\right) = R_{nl}\left(r\right)Y_{l}^m\left(\bm{\Omega}\right)\chi_{\sigma}
\end{eqnarray}
where $m$ labels the $z$-component orbital angular momentum and $\chi_{\sigma}$ describes the spin-$\sigma$ state. For the present study, $p$ orbitals with $l=1$ will be under consideration.
The matrix elements of the Coulomb interaction for spin-$1/2$ electrons in a fixed $nl$ atomic shell are given by
\begin{eqnarray}
	U_{m_1\sigma_1,m_2\sigma_2;m_3\sigma_3,m_4\sigma_4}&=&\int d\bm{r}d\bm{r}^\prime
	\phi^*_{m_1\sigma_1}\left(\bm{r}\right)
	\phi^*_{m_2\sigma_2}\left(\bm{r}^\prime\right)
	\frac{1}{|\bm{r}-\bm{r}^\prime|}
	\phi_{m_3\sigma_3}\left(\bm{r}^\prime\right)	
	\phi_{m_4\sigma_4}\left(\bm{r}\right).	
\end{eqnarray}
Making use of the multipole expansion
\begin{eqnarray}
	\frac{1}{|\bm{r}-\bm{r}^\prime|} 
	= \sum_{km} \frac{4\pi}{2k+1}\frac{r_{<}^k}{r_{>}^{k+1}} 
	Y_k^m\left(\bm{\Omega}\right)
	Y_k^{m*}\left(\bm{\Omega}^\prime\right)
\end{eqnarray}
with $r_{<}$ and $r_{>}$ denoting the smaller and larger modules of $\bm{r}$ and $\bm{r}^\prime$ vectors respectively,
the interaction matrix elements can be further expressed as
\begin{eqnarray}
	U_{m_1\sigma_1,m_2\sigma_2;m_3\sigma_3,m_4\sigma_4} &=& \sum_{k} \delta_{m_1+m_2,m_3+m_4}F^k G_k\left(m_1,m_4\right)G_k\left(m3,m2\right)
	\delta_{\sigma_1\sigma_4}\delta_{\sigma_2\sigma_3}.
	\label{eq:IME-S}
\end{eqnarray}
In the above equation, the Slater-Condon-Shortley parameters~\cite{Slater1929,Condon1931}
\begin{eqnarray}
	F^k=\int drdr^\prime r^2r^{\prime 2} |R_{nl}|^2\left(r\right)|R_{nl}|^2\left(r^\prime\right)
\end{eqnarray}
involve the integral of radial functions, and the Gaunt coefficients~\cite{Gaunt1929}
\begin{eqnarray}
	G^k\left(m,m^\prime\right)&=&\left(-1\right)^m\left(2l+1\right)
	\left(
	\begin{matrix}
		l & l & k \\
		0 & 0 & 0
	\end{matrix}
	\right)
	\left(
	\begin{matrix}
		l & l & k \\
		m^\prime & -m & m-m^\prime
	\end{matrix}
	\right) 
\end{eqnarray}
can be expressed in terms of Wigner $3j$ symbols
$\left(
	\begin{matrix}
		j_1 & j_2 & j_3 \\
		m_1 & m_2 & m_3
	\end{matrix}
\right)$.
It is noteworthy that the interaction matrix elements in Eq.~(\ref{eq:IME-S}) parametrized in terms of independent Slater-Condon-Shortley parameters $F^k$ respect the full orbital rotation symmetry. 

The matrix elements of Coulomb interaction in $\bm{p}=\left(p_x,p_y,p_z\right)$ orbital basis are given by
\begin{eqnarray}
	V\left(\mu_1\sigma_1,\mu_2\sigma_2;\mu_3\sigma_3,\mu_4\sigma_4\right)
	=\sum_{\{m\}}T^*_{\mu_1 m_1} T^*_{\mu_2 m_2}
	U_{m_1\sigma_1,m_2\sigma_2;m_3\sigma_3,m_4\sigma_4}
	T_{\mu_3 m_3} T_{\mu_4 m_4}
\end{eqnarray}
with aid of the transformation $\bm{Z}=T\bm{Y}$ from spherical to tesseral harmonics
\begin{subequations}
	\begin{eqnarray}
	Z_{11}^\text{c}&=&\sqrt{\frac{3}{4\pi}}\frac{x}{r}
	=\frac{1}{\sqrt{2}}\left[Y_1^{-1}\left(\bm{\Omega}\right)-Y_1^1\left(\bm{\Omega}\right)\right],\\
	Z_{11}^\text{s}&=&\sqrt{\frac{3}{4\pi}}\frac{y}{r}
	=\frac{i}{\sqrt{2}}\left[Y_1^{-1}\left(\bm{\Omega}\right)+Y_1^1\left(\bm{\Omega}\right)\right],\\
	Z_{10}  &=&\sqrt{\frac{3}{4\pi}}\frac{z}{r}=Y_1^0\left(\bm{\Omega}\right).
	\end{eqnarray}
\end{subequations}
The calculations of Coulomb interaction can be further simplified by taking the advantage of Fermi statistics. We therefore rewrite the matrix elements  
\begin{eqnarray}
	I\left(\mu_1\sigma_1,\mu_2\sigma_2;\mu_3\sigma_3,\mu_4\sigma_4\right)
	=V\left(\mu_1\sigma_1,\mu_2\sigma_2;\mu_3\sigma_3,\mu_4\sigma_4\right)
	-V\left(\mu_1\sigma_1,\mu_2\sigma_2;\mu_4\sigma_4,\mu_3\sigma_3\right).
\end{eqnarray}
The former denotes the direct term, while the latter indicates the exchange term.
We skip the tedious calculation of the Coulomb interaction matrix elements and here only summarise the non-vanish elements in Table~\ref{tab:Slater-S}. Collecting these elements, the multiorbital Hubbard interaction reads
\begin{eqnarray}
	H_{\text{I}} 
	=\left(F^0+\frac{4}{25}F^2\right)\sum_{\mu}\hat{n}_{\mu\uparrow}\hat{n}_{\mu\downarrow}
	+\left(F^0-\frac{7}{50}F^2\right)\sum_{\mu<\nu}\hat{n}_{\mu}\hat{n}_{\nu}
	+\frac{3}{25}F^2\sum_{\mu\ne\nu}
	\left(p_{\mu\uparrow}^\dagger p_{\mu\downarrow}^\dagger p_{\nu\downarrow} p_{\nu\uparrow}
	- \bm{S}_{\mu}\cdot\bm{S}_{\nu} \right),
\end{eqnarray}
which recovers the Kanamori form of $t_{2g}$ orbitals~\cite{Kanamori1963} but with the interaction parameters being {\it ab  initio} determined from $p$ orbitals.

\end{document}